\documentclass[12pt]{article}

\usepackage{amsmath,amssymb,amsfonts}
\usepackage[left=30mm,right=30mm,top=30mm,bottom=30mm]{geometry}
\usepackage{titlesec} 
\usepackage{setspace} 
\usepackage{textcomp} 
\usepackage{hyperref}
\usepackage{color}
\usepackage{graphicx,epsfig}
\usepackage{epstopdf} 
\usepackage{verbatim} 

\titleformat{\section}
  {\large\bfseries}{\thesection}{0.5em}{}
\titleformat{\subsection}
  {\normalfont\bfseries}{\thesubsection}{0.5em}{}

\makeatletter
\renewcommand{\fnum@figure}{\textbf{Fig.~\thefigure}}
\renewcommand{\fnum@table}{\textbf{Table~\thetable}}
\makeatother

\numberwithin{equation}{section}

\allowdisplaybreaks[1]

\DeclareMathOperator{\tr}{Tr}

\def\al{\alpha} \def\be{\beta} \def\ga{\gamma} \def\Ga{\Gamma}
\def\de{\delta} \def\De{\Delta} \def\ep{\epsilon} 
\def\th{\theta}  \def\ka{\kappa} 
\def\La{\Lambda} \def\si{\sigma}  
  \def\om{\omega} \def\Om{\Omega}

\def\bga{{\bar \gamma}} \def\bka{{\bar \kappa}}

 \def\tY{\widetilde Y}

\def\cH{\mathcal H} \def\tcH{\widetilde{\mathcal H}}

\def\cL{\mathcal L}

\def\ui{{r}} \def\uj{{s}} \def\uk{{t}} \def\ul{{u}}

\def\p{\partial}

\def\dk{\delta_\kappa}

\def\nn{\nonumber \\}
\def\<{\langle}
\def\>{\rangle}

\begin{document}

\begin{titlepage}

\renewcommand{\thefootnote}{\fnsymbol{footnote}}

\rightline{APCTP-Pre2009-012}

\vspace{15mm} \baselineskip 9mm
\begin{center}
{\Large\bf Heterotic Action in SUGRA-SYM Background}
\end{center}

\baselineskip 6mm \vspace{7mm}
\begin{center}
{\large Jaemo Park}$^{1,2}$\footnote{\tt jaemo@postech.ac.kr},
{\large Cheol Ryou}$^{1}$\footnote{\tt string@postech.ac.kr} {and}
{\large Woojoo Sim}$^{3}$\footnote{\tt space@apctp.org}
\\[5mm]

{\small\it
$^1$Department of Physics, POSTECH, \\
Pohang 790-784, Korea \\ \vspace{0.3cm}

$^2$Postech Center for Theoretical Physics (PCTP), POSTECH, \\
Pohang 790-784, Korea \\ \vspace{0.3cm}

$^3$Asia Pacific Center for Theoretical Physics, \\
Pohang 790-784, Korea \\ \vspace{0.3cm}
}
\end{center}

\vspace{5mm}
\begin{center}
{\bf Abstract}
\end{center}
\indent We consider the generalization of the heterotic action
considered by Cherkis and Schwarz where the chiral bosons are
introduced in a manifestly covariant way using an auxiliary
field. In particular, we construct the kappa-symmetric heterotic
action in ten-dimensional supergravity background coupled to super
Yang-Mills theory and prove its kappa-symmetry. The usual Bianchi
identity of Type I supergravity with super Yang-Mills $dH_3= -\tr
F\wedge F $ is crucially used. For technical reason, the
Yang-Mills field is restricted to be abelian.

\end{titlepage}

\renewcommand{\thefootnote}{\arabic{footnote}}
\setcounter{footnote}{0}

\section{Introduction}

In \cite{ps}, supersymmetric heterotic action with kappa-symmetry
in flat space is constructed motivated by the bosonic truncation
of the heterotic action obtained by the M5 brane action wrapping
K3 \cite{cherkis}. In the construction of \cite{ps, cherkis}, the
realization of the degrees of freedom associated with the current
algebra is novel, which is generalization of PST approach of the
realization of chiral two-form in M5-brane \cite{pasti}.
Thus it would be interesting to work out various aspects of
this new heterotic action with kappa-symmetry.
As one such attempt we generalize \cite{ps} to the kappa-symmetric
heterotic action in arbitrary background of Type I coupled to
super Yang-Mills. Due to the technical subtleties,
we consider only abelian gauge fields.
The form of the heterotic action can be guessed from the M5 brane
action wrapping on K3. Once written, it is straightforward to work
out the kappa-symmetry of such action. One point is worthy of
mention. Along the proof of the kappa-symmetry the usual relation
of the Type I supergravity coupled to super Yang-Mills
\begin{align}\label{dh3}
dH_3= -\tr F\wedge F
\end{align}
is crucially used. Similar feature appears in the kappa-symmetric
action of the membrane with the Horava-Witten boundary worked out
by Cederwall \cite{cederwall}.

In the current formalism of the heterotic string theory, it is
easy to couple the abelian gauge field to the worldsheet of the
heterotic string but the non-abelian generalization is not
straightforward. Related fact is that if we wrap M5 brane on K3,
we assume that involved K3 is smooth so that we can carry out the
usual Kaluza-Klein compactification so that we obtain the
heterotic string in 2-dimensions. Thus the non-abelian
generalization is related to figuring out the additional degrees
of freedom we should keep when we try to wrap M5 brane on singular
K3. This is also related to working out the full chiral current
algebra of the heterotic string when we assume the gauge group
$E_8 \times E_8$ or $SO(32)$ in the current formalism. Thus if we
assume that the heterotic string realizes $E_8 \times E_8$ or
$SO(32)$ current algebra, abelian gauge field configuration we are
considering is a subclass of the general field configuration.  In
the proof of the kappa symmetry, the background equation of motion
we obtain has straightforward generalization to non-abelian gauge
field background so we can guess the general form of the
background equation of motion once we figure out the correct
coupling of the general gauge field configurations to the world
sheet of the heterotic string. This is an interesting topic per
se, but it's beyond the scope of the current letter.

In the below we  assign coordinate indices as follows:
\begin{itemize}\setlength\itemsep{-\parsep}
\item 10-d and 11-d target (tangent) : $Z^A=(X^a,\th^\al)$.
\item 10-d and 11-d target (curved) : $Z^M=(X^m,\th^\mu)$.
\item 2-d and 6-d worldvolume : $\si^i$, $\si^j$, $\cdots$.
\item K3 : $X^\ui=\si^\ui$, $X^\uj=\si^\uj$, $\cdots$ (static gauge).
\end{itemize}

\section{Heterotic string in SUGRA-SYM background from M5-brane}
\label{sec:m5}

In \cite{ps}, a 10-d supersymmetric heterotic action in flat
background has been constructed from the M5-brane action doubly reduced
on K3 surface. Here we extend the 10-d action to have the
SUGRA-SYM background. As in the flat case, the action with
generic background can be guessed from the M5-brane action with
11-d SUGRA background. Then, we can justify the action by working
out the gauge symmetry and the kappa symmetry as given in
section~\ref{sec:kappa}.

The M5-brane action in 11-d supergravity background is given as
\cite{bandos,bandos2}
\begin{align} \label{M5_action}
\cL_1 &= -\sqrt{ -\det \biggl( G_{ij}
            + i\frac{\tcH_{ij}}{\sqrt{-Gu^2}} \biggr)}, \nn
\cL_2 &= -\frac{1}{4u^2} \tcH^{ij}\cH_{ijk}G^{kl}u_l, \nn
S_{WZ} &=  \int \biggl( c_6 + \frac{1}{2}\cH_3\wedge c_3 \biggr).
\end{align}
Here the worldvolume fields are $Z^M$, $A_2$
and auxiliary scalar $a$ ($u_i=\p_i a$) which makes the action
covariant.
In the action, $\cH_3$ is the supersymmetrized field strength of
$A_2$:
\begin{align}
\cH_3 = dA_2-c_3, \quad \tcH^{\mu\nu} =
\frac{1}{6}\ep^{ijklmn} \cH_{klm} u_n,
\end{align}
and $\Pi_i^A$, $c_3$, $c_6$ are the pullbacks of
the supervielbein and the superforms of the background:
\begin{align} \label{pb_gen}
\Pi_i^A &= \p_i Z^M E_M\!^A, \nn
c_{ijk} &= \p_i Z^M \p_j Z^N \p_k Z^P C_{PNM}, \nn
G_{ij} &= \Pi_i^A \Pi_j^B  \eta_{AB}.
\end{align}

From the above action, we can infer the 10-d heterotic action in
SUGRA-SYM background. This has been done for a simpler case in
\cite{ps,cherkis} by wrapping the M5-brane on K3 surface to
get a 7-d action and then by lifting it  to 10-d. Here we adopt
the same scheme but do not perform the reduction in a precise way.
Instead, we guess the reduction and the 10-d action from the past
experience of the calculations in \cite{ps}.

First, the worldvolume fields in eq.~\eqref{M5_action} can be
reduced to the worldsheet fields as
\begin{align} \label{red}
(dA_2)_{ijk} &\to \p_i Y^I b_{I\ui\uj}, \nn
c_{ijk} &\to -A_i^I b_{I\ui\uj}, \nn
c_{i_1 \cdots i_6} &\to -B_{ij}\om_{\ui\uj\uk\ul},
\end{align}
where $b_I$ and $\om$ are the harmonic 2-forms and
the volume form on K3, respectively.
For instance, the reduction of $c_3$ can be inferred as
\begin{align}
c_{irs}=\p_i Z^M \p_\ui Z^N \p_\uj Z^P C_{PNM}
    \to -\p_i Z^M A_M^I b_{I\ui\uj} = -A_i^I b_{I\ui\uj}.
\end{align}

Note that among 22 harmonic 2-forms on K3,
19 are anti-self-dual and 3 are self-dual,
which yields the correct degrees of freedom of chiral
bosons $Y^I$ (upon gauge fixing) for
the heterotic string action compactified on $T^3$.

In eq.~\eqref{red}, as the action is lifted up to 10-d,
$A_i$ and $B_{ij}$ are understood as the pullbacks of 10-d superforms:
\begin{align}
A_i^I &= \p_i Z^M A_M^I, \nn
B_{ij} &= \p_i Z^M \p_j Z^N B_{NM}.
\end{align}

Then, $\cH_3=dA_2-c_3$ reduces to
\begin{align} \label{red_h}
\cH_{ijk} &\to (\p_i Y^I + A_i^I) b_{I\ui\uj}
    = D_i Y^I b_{I\ui\uj}, \nn
\tcH_{ij} &\to  \tY^I b_{I\ui\uj},
\end{align}
where $\tY^I=\ep^{ij}D_i Y^I u_j$ and $D_i$ is a gauge covariant
derivative given that
\begin{align}
\de A_i^I = \p_i \La^I, \quad \de Y^I = -\La^I.
\end{align}
Note that in this way we obtain the coupling of the abelian gauge
field to the heterotic string.

In eq.~\eqref{red}, the harmonic 2-forms $b_I$ contribute
to the reduced action in terms of
\begin{align} \label{LM}
L_{IJ} = \int_{K3} b_I\wedge b_J, \quad M_{IJ} = \int_{K3}
b_I\wedge *b_J.
\end{align}
Here, $L$ is the intersection matrix of the 2-cycles on K3:
\begin{align}
L = -\Ga_8\oplus -\Ga_8\oplus \si\oplus\si\oplus\si,
\end{align}
where $\Ga_8$ is the same as the Cartan matrix of $E_8$ and
$\si=\big(\begin{smallmatrix} 0&1 \\ 1&0 \end{smallmatrix} \big)$.
Then, as lifted to 10-d, L can be interpreted as the Cartan
matrix of $E_8\times E_8$ or $SO(32)$, which is one of the two
gauge groups of 10-d heterotic theory. (Note, in 10-d, $L=-M$.)
Note that at the tree level, the possible gauge groups are not
determined. Only if we consider one-loop consistency condition,
the possible gauge groups can be fixed.

In this sense, $Y^I$ in eq.~\eqref{red} are understood as the
scalars which represent the 16 bosonic degree of freedom
compactified on $T^{16}$ in 10-d heterotic theory, yielding the
Narain lattice which is the same as the root lattice of $E_8\times
E_8$ due to the one-loop modular invariance. In addition, the
gauge field $A^I$, which couples to $L_{IJ}$, is an element of the
Cartan subalgebra of $E_8\times E_8$, so as to be abelian
($F=dA$), being consistent with the form of the action given
below. We can also take $L = -\Ga_{16}$, root lattice of $SO(32)$
gauge group for $SO(32)$ heterotic string theory~\cite{cherkis}.

Finally, taking above field reductions into account, we propose
the heterotic action whose background is 10-d SUGRA-SYM:
\begin{align} \label{het_action}
\cL_1 &= -\sqrt{-G}\sqrt{1 - \frac{\tY^I L_{IJ}\tY^J}{Gu^2}
    +\biggl(\frac{\tY^I L_{IJ}\tY^J}{2Gu^2}\biggr)^2}, \nn
\cL_2 &= -\frac{\tY^I L_{IJ}D_i Y^J u^i}{2u^2}, \nn
\cL_{WZ} &= -\frac{1}{4}\ep^{ij} Y^I L_{IJ}F_{ij}^J
    -\frac{1}{2}\ep^{ij} B_{ij}.
\end{align}
This action is invariant under gauge transformations
\begin{align} \label{gauge}
\de A_i^I = \p_i \La^I, \quad \de Y^I = -\La^I,
\quad \de B_{ij} = \frac{1}{2} \La^I L_{IJ} F_{ij}^J.
\end{align}

\section{Kappa symmetry of 10-d heterotic action}
\label{sec:kappa}

Now let us show the kappa invariance of the action. First, we
propose the following kappa variations of the worldsheet fields in
the heterotic action.
\begin{align}\label{kappa}
\dk Z^M &= \De^\al E_\al\!^M \quad
(\dk Z^M E_M\!^\al = \De^\al, \quad \dk Z^M E_M\!^a = 0), \nn
\dk Y^I &= K^I, \nn
\dk A_i^I &= -\p_i K^I + F_i^I, \nn
\dk B_{ij} &= -\De^\al \Pi_i^A \Pi_j^B (dB)_{\al AB}
    +\textrm{ total derivative},
\end{align}
where $\De^\al=\bka^\be (1-\Ga)_\be^\al$ and
\begin{align}
K^I &= -\dk Z^M A_M^I = -\De^\al E_\al\!^M A_M^I, \nn
F_i^I &= \dk Z^M \p_i Z^N F_{MN}^I = \De^\al \Pi_i^A F_{\al A}.
\end{align}
Then, the variations of related fields in the action are evaluated as
\begin{align}\label{kappa_2}
\dk\Pi_i^a &= \De^\al \Pi_i^B (T_{\al B}\!^a-\Om_{\al B}\!^a), \nn
\dk G_{ij} &= \De^\al T_{\al A}\!^a \Pi_{(i}^A \Pi_{j)}^b \eta_{ab}, \nn
\dk (D_i Y^I) &= F_i^I, \nn
\dk \tY^I &= \ep^{ij}F_i^I u_j,
\end{align}
where $T$ and $\Om$ are the torsion and the connection one-form of the background superspace, respectively.

Note here that the kappa variation of $Y^I$ in eq.~\eqref{kappa}
can be inferred from the variation of $A_2$ in M5-brane
side:
\begin{align}
\dk A_2 &= -\dk Z^M dZ^N dZ^P C_{PNM} \nn
\to \dk Y^I b_{I\ui\uj} &= -\dk Z^M A_M^I b_{I\ui\uj},
\end{align}
yielding
\begin{align}
\dk Y^I = -\dk Z^M A_M^I = -\De^\al E_\al\!^M A_M^I \equiv K^I.
\end{align}

For the details of the kappa variations of the fields,
see appendix~\ref{sec:app_a}.

\subsection{Kappa Invariance: scheme}
The method we use in the proof of the kappa symmetry
is similar to the one used in \cite{ps}
for the flat case without $A_i$.
We organize terms in $\dk \cL$ in the order of
$\tY^I=\ep^{ij}D_i Y^I u_j$ and check the invariance order by order.
This means that all terms in the variation should be recast
in terms of the powers of $\tY$.
For the case of $\dk \cL_1$, this is evident since all terms in $\cL_1$
are already written in terms of $\tY$.

Then we can expect the terms in $\dk(\cL_2+\cL_{WZ})$ to be
recast in terms of $\tY$ as well,
considering the Nambu-Goto type action we are dealing with here.
Indeed, as advertised in the introduction,
this can be realized in the aid of the $H_3$
in eq.~\eqref{dh3} and \eqref{h3}.
For the details, see section~\ref{sec:proof}.

In the proof, $U$ and $T$ are defined as
\begin{align}\label{ut}
4\De U = \dk (\cL_1)^2, \quad
2\De T = \dk \cL_2 + \dk \cL_{WZ},
\end{align}
where $\De=\bar{\ka}(1-\Ga)$ and we have suppressed
the spinor indices.
Then, we can take a quantity $\rho$ satisfying
\begin{align} \label{pf1}
U = \rho T, \quad \rho^2 = (\cL_1)^2,
\end{align}
which ensures the kappa invariance taking $\Ga=\rho/\cL_1$:
\begin{align} \label{pf2}
\dk \cL &= 2\De \bigg( \frac{U}{\cL_1} + T \bigg)
= 2\bka(1 - \Ga)(1 + \Ga) T =0.
\end{align}
The details of the proof will be given in section~\ref{sec:proof}.

In the proof of kappa symmetry, we need to consider the on-shell
constraints on the SUGRA-SYM backgrounds. Using the constraints,
we will see that the proof goes in the same way as the flat case with
the addition of gauge field $A_i$. Therefore, let us first look through the
SUGRA-SYM constraints.

\subsection{SUGRA-SYM constraints}
\label{sec:sugra}
The on-shell constraints on the 10-d heterotic background read
\cite{kallosh,witten,candiello}
\begin{alignat}{2}\label{const}
T_{\al\be}\!^a &=2(\Ga^a)_{\al\be},
    &\quad T_{\al b}\!^a &= 0, \nn
H_{\al\be a} &=2(\Ga_a)_{\al\be},
    & H_{\al ba} &= H_{\al\be\ga} =0, \nn
F_{a \al} &=2(\Ga_a)_{\al\be}\chi^\be,
    & F_{\al\be} &=0.
\end{alignat}
Here, $H_3$ is written as
\begin{align}\label{h3}
H_3 = dB_2 - \frac{1}{2}A^I\wedge F^J L_{IJ}
= dB_2 - \tr (A\wedge F)
\end{align}
Note that $H_3$ is a gauge invariant field strength,
which can be seen from eq.~\eqref{gauge}.
In addition, reminding that $A$ is an element of
the Cartan subalgebra of $E_8 \times E_8$ and that
$L_{IJ}$ is the Cartan matrix of the algebra,
we see that eq.~\eqref{h3} is satisfied since
\begin{align}
\tr (A\wedge F) = A^I\wedge F^J \tr(h_I h_J)
= \frac{1}{2}A^I\wedge F^J L_{IJ}.
\end{align}
for a suitable choice of the basis $h_I$ of the Cartan subalgebra.

Using the constraints in eq.~\eqref{const},
the kappa variations of related fields reduce as
\begin{align}\label{kappa1}
\dk G_{ij} &= \De^\al T_{\al A}\!^a \Pi_{(i}^A \Pi_{j)a}
    = 2\De \ga_{(i} \Pi_{j)}, \nn
\dk(D_i Y^I) &= F_i^I = \De^\al \Pi_i^A F_{\al A}^I
    = -2\De \ga_i \chi^I,
\end{align}
and, up to a total derivative,
\begin{align}\label{kappa2}
\dk B_{ij} &= -\De^\al \Pi_i^A \Pi_j^B (dB)_{\al AB} \nn
    &= -\De^\al \Pi_i^A \Pi_j^B
        \big(H_{\al AB}+ \tr (A\wedge F)_{\al AB} \big) \nn
    &= 2\De \ga_{[i}\Pi_{j]}
        - \frac{1}{2}\De^\al \Pi_i^A \Pi_j^B
    L_{IJ}(A^I\wedge F^J)_{\al AB}.
\end{align}
with $\ga_i = \Pi_i^a \Ga_a$ in the final expression of each
formula. (We have not suppressed the spinor indices for the
3-superforms to be explicit.)  Here the beauty lies in the fact
that the second term of $\dk B_{ij}$ exactly cancels some terms
that appears in $\dk(\cL_2+\cL_{WZ})$, which is needed to make
every term of $T$ written in terms of the powers of $\tY$. (See
the detailed derivations given below.)

Note in eq.~\eqref{kappa1} that $\dk G_{ij}$ has the same form as the flat
case~\cite{ps} where $\Pi_j$ reduces to $\p_j\th$. Likewise, the
first term of $\dk B_{ij}$ has the same form as the flat case.
Consequently, as mentioned above, this implies that the proof works in
the same way as the flat case. However, in \cite{ps} the induced
gauge field $A_i$ was not considered. Therefore, while we
describe the whole proof in the followings, we concentrate on the
part where the gauge field has an effect on.

\subsection{Details of the proof}
\label{sec:proof}
Now let us complete the proof by showing $U=\rho T$ with
$\rho^2 =(\cL_1)^2$.

First, $\rho$ can be determined as
\begin{align}
\rho = \bga - \frac{\tY L\tY}{2Gu^2} \bga.
\end{align}

Then, we will show $U = \rho T$ order by order in $\tY$:
\begin{align}\label{order}
U_0 &= \rho_0 T_0, \nn
U_1 &= \rho_0 T_1, \nn
U_2 &= \rho_0 T_2 + \rho_2 T_0, \nn
U_3 &= \rho_2 T_1,\nn
U_4 &= \rho_2 T_2,
\end{align}
where the subscripts denote the order of $\tY$ in the terms
and we have used $\rho_1=0$.

Here $U$ and $T$ are determined from eq.~\eqref{ut}
using the kappa variations of
the worldsheet fields in eq.~\eqref{kappa1} and \eqref{kappa2}
where SUGRA-SYM constraints have been considered.
First, from $\dk (\cL_1)^2$ we get $U$ as
\begin{align}\label{u}
U_0 &= -G\ga^i\Pi_i, \nn
U_1 &= \frac{\tY L}{u^2}\ep^{ij} u_i \ga_j \chi, \nn
U_2 &= \frac{\tY L\tY}{(u^2)^2} u^i u^j \ga_i \Pi_j, \nn
U_3 &= -\frac{\tY L\tY}{2G(u^2)^2}
    \tY L \ep^{ij}u_i \ga_j \chi, \nn
U_4 &= -\frac{(\tY L\tY)^2}{4G(u^2)^3}
    (2\ga^i u^j-\ga^j u^i)u_i \Pi_j.
\end{align}
Next, $T$ is determined by $\dk\cL_2$ and $\dk\cL_{WZ}$:
\begin{align}
\dk \cL_2 &= -\frac{1}{2u^2}\tY L F_i u^i
-\frac{\ep^{ij}}{2u^2}F_i u_j LD_k Y u^k
+\frac{\tY L \tY}{G(u^2)^2}
    \ep^{ik}u_k u^j \De\ga_{(i} \Pi_{j)}, \nn
\dk \cL_{WZ} &=
\frac{\ep^{ij}}{2}\big(K L \p_i A_j + F_i L \p_j Y \big)
-2\ep^{ij}\De\ga_i\Pi_j
-\frac{\ep^{ij}}{4}\De^\al \Pi_i^A \Pi_j^B (A\wedge F)_{\al AB},
\end{align}
where we have used the identity
\begin{align}\label{id_1}
\ep^{ij}\ep_{kl}= - \de^i_k \de^j_l + \de^i_l \de^j_k.
\end{align}
Here, adding the first two terms of $\dk\cL_2$
to the first two terms of $\dk\cL_{WZ}$
and using eq.~\eqref{id_1} yield
\begin{align}\label{t_1}
-\frac{1}{u^2}\tY L F_i u^i
+\frac{\ep^{ij}}{2}\big(K L \p_i A_j - F_i L A_j \big).
\end{align}
Note that the last two terms in eq.~\eqref{t_1} are not written
in terms of $\tY$ and we can expect it to be canceled out
with some other term in the variation.
In fact, those two terms are recast as
\begin{align}
-\frac{\ep^{ij}}{4}\De^\al \Pi_i^A \Pi_j^B
    L_{IJ}(A^I\wedge F^J)_{\al AB},
\end{align}
which exactly cancels out the last term of $\dk\cL_{WZ}$, making
all terms in $\dk(\cL_2+\cL_{WZ})$ written in terms of $\tY$. With
eq.~\eqref{ut}, we are led to
\begin{align}\label{t}
T_0 &= -\ep^{ij}\ga_i\Pi_j, \nn
T_1 &= \frac{\tY L}{u^2} u^i \ga_i \chi, \nn
T_2 &= \frac{\tY L \tY}{2G(u^2)^2}
    \ep^{ik}u_k u^j \ga_{(i} \Pi_{j)}.
\end{align}

Now, we are left with the verification of $U=\rho T$.
However, we have seen in section~\ref{sec:sugra}
that the kappa variations of the worldsheet fields have the same forms,
the SUGRA-SYM constraints being considered, as the ones in the flat background.
Then, since in \cite{ps} the kappa symmetry for the flat
case without gauge field $A$ has been proved,
it is sufficient here to consider only the terms related with
the variation of $A_i$ in showing $U=\rho T$.

Considering the kappa variations of the fields in
eq.~\eqref{kappa1} and
\eqref{kappa2}, we notice that only $\dk D_i Y$ is related to $\dk
A_i$ and that the terms including $\dk (D_i Y)$ in $U=\rho T$ are the
1st and the 3rd order terms. As a result, we only need to show
$U=\rho T$ in the odd orders \footnote{For the proof of $U=\rho T$
in even orders, refer to the appendix of \cite{ps}.}.

However, note that $U_3$ can be recast as,
for $[\rho_0,\rho_2]=0$,
\begin{align}
U_3=\rho_2 \rho_0^{-1} U_1,
\end{align}
and that if $U_1=\rho_0 T_1$ is satisfied,
$U_3=\rho_2 T_1$ is automatic.
Therefore, we only need to show $U_1=\rho_0 T_1$
to complete the proof.
Then, using $\rho_0=\bga$,
\begin{align}
\rho_0 T_1 = \frac{\tY L}{u^2} \bga\ga^i u_i \chi
    = \frac{\tY L}{u^2} \ep^{ij} u_i \ga_j \chi = U_1,
\end{align}
where we have used the identity $\bga\ga^i=\ep^{ij}\ga_j$.

Thus, we have completed the proof of the kappa invariance.

\medskip
\subsection*{Acknowledgments}
We thank Masayoshi Yamamoto for the collaboration at the initial
stage of the project. We are especially indebted to him for the
material presented in the section 2 of the main text. J.P. is
supported in part by KOSEF Grant R01-2008-000-20370-0, by the
National Research Foundation of Korea(NRF) grant funded by the
Korea government(MEST) with the grant number 2005-0049409 and the
grant number  2009-0085995.

\medskip
\section*{Appendix}
\appendix

\section{Kappa Variations} \label{sec:app_a}
The kappa variations of the pullback $A_i^I$ is
evaluated as
\begin{align}
\dk A_i^I &= \dk (\p_i Z^M A_M^I) \nn
    &= \p_i \dk Z^M A_M^I + \p_i Z^M \dk A_M^I \nn
    &= \p_i (\dk Z^M A_M^I) - \dk Z^M \p_i A_M^I + \p_i Z^M \dk
    A_M^I \nn
    &= \p_i (\dk Z^M A_M^I) - \dk Z^M \p_i Z^N \p_N A_M^I +
     \p_i Z^M \dk Z^N \p_N A_M^I \nn
    &= -\p_i K^I + F_i^I,
\end{align}
where, for $F^I=dA^I$,
\begin{align}
F_i^I = \dk Z^M \p_i Z^N F_{MN}^I = \De^\al \Pi_i^A F_{\al A}.
\end{align}

Note, from $\dk Y^I$ and $\dk A_i^I$ above,
\begin{align}
\dk (D_i Y^I) &= \p_i \dk Y^I + \dk A_i^I = F_i^I, \nn
\dk \tY^I &= \ep^{ij}F_i^I u_j,
\end{align}
which is crucial for the kappa symmetry in the sense that
$\De$ can be factorized as an overall factor.

The kappa variation of $B_{ij}$
is evaluated similarly to $\dk A_i$, up to a total derivative:
\begin{align}
\dk B_{ij} = -\dk Z^M \p_i Z^N \p_j Z^P (dB)_{MNP}
    = -\De^\al \Pi_i^A \Pi_j^B (dB)_{\al AB}.
\end{align}

Finally, we need to evaluate $\dk\Pi_i^A$:
\begin{align}
\dk\Pi_i^A &= \dk(\p_i Z^M E_M\!^A) \nn
&= \p_i \dk Z^M E_M\!^A + \p_i Z^M \dk E_M\!^A \nn
&= \p_i (\dk Z^M E_M\!^A) - \dk Z^M \p_i E_M\!^A
    + \p_i Z^M \dk E_M\!^A \nn
&= \p_i \De^A + \De^\al \Pi_i^B (T_{\al B}\!^A-\Om_{\al B}\!^A),
\end{align}
where we have used $\De^a \equiv 0$ (as a notational convention),
which is consistent with
\begin{align}
\dk Z^M E_M\!^\al = \De^\al, \quad \dk Z^M E_M\!^a = 0.
\end{align}
As a result,
\begin{align}
\dk\Pi_i^a = \De^\al \Pi_i^B (T_{\al B}\!^a - \Om_{\al B}\!^a),
\end{align}
which yields
\begin{align}
\dk G_{ij} &= \dk\Pi_{(i}^a\Pi_{j)}^b \eta_{ab} \nn
&= \De^\al \Pi_{(i}^B \Pi_{j)}^b (T_{\al B}\!^a - \Om_{\al B}\!^a) \eta_{ab} \nn
&= \De^\al T_{\al B}\!^a \Pi_{(i}^B \Pi_{j)}^b \eta_{ab},
\end{align}
Note here that the connection $\Om$ has not contributed
since $\Om_{\al B b} + \Om_{\al b B} =0$.

\end{document}